# A secured communication link design using narrow line of sight technique

Swapan Bhattachrya1[a] , Devmalya Banerjee 2[b], Abirlal Biswas 3[b], Pijush Biswas 4[a],

1a  Asansol Engineering College, Asansol-04, India
2b Asansol Engineering College, Asansol-04, India
3b Dept. of Computer Sc. And Engg.,C.U., Kolkata – 09, India
4a A. K. Choudhury School of I.T., C.U., Kolkata-09 , India

pijushindia@gmail.com

**Abstract:** In recent wireless networking optical communication is an emerging field. Another novel approach of optical communication using Narrow Line of Sight technique is discussed. Array of trans-receivers are placed on the ceiling to communicate with the mobile nodes (users / trans-receivers) instead of a single transmitter. An intelligent position tracking technique based on the informations available from prepositioned sensors and using them in simple geometric equations is described. Intelligent cell concept is used to facilitate the tracking system. The whole wide area is divided into several sub areas and some of the sub areas are designated as boundary sub areas which hold the data regarding the movement of the mobile nodes. The Diffie-Hellman simple authenticated key agreement is used for secure communication.

## 1. INTRODUCTION

In the emerging field of computing, wireless networking is one of the most attractive trends. Users want to establish a high speed communication link among themselves as well as with the external environment without being tethered to wires as well as wall jacks. So mobility has become an important issue in common office networking environments. At present there are two parallel important technologies, i.e. the radio communication link and the wireless optical link. [1] The wireless optical link has its own advantage like high bandwidth, security and low component price. So with the advent of new technologies the wireless optical network becomes an attractive alternative to its nearest competitor. Development of several optical network architecture has been accomplished according to two main categories : the diffused system and Line of Sight (LOS) system. [2,3] The diffused system suffers from multi path effect. The achievable bit rate is about 52 Mbps in a 5 meter room [4].

Again  the LOS system is only limited by the capability of the component used. With the rapid advancement of transmitter and receiver designing techniques, it gradually becomes a suitable choice.

The Line of  Sight can be categorized in Wide Line of Sight (WLOS) and Narrow Line of Sight (NLOS) categories. In WLOS one ceiling mounted transmitter is used to cover a wide area with its spread beam of light. As a result the launch power of the laser transmitter should be relatively high resulting a threat to eye safety.

The paper concentrates on the NLOS system in which the spread beam from the laser diode is directed towards the optical receiver. The covered area being small the launch power from the transmitter will be within the safety limit. The team at the British Telecomm Laboratory first suggested that with an associated tracking mechanism, mobility of a user can be achieved. D.Wisely *et al* and  A.M.Street *et al* proposed a NLOS system based on the used of an array of vertical cavity surface emitting lasers ( VCSEL). This paper proposes a total compact framework of a multiuser secured communication link designed along with a simple position tracking method.

## 2.  PROPOSED ARCHITECHTURE

In WLOS technique a ceiling mounted transmitter communicates with the user. As suggested by Bellon *et al* [5] the single laser source will be replaced by an array of VCSELs. Thus the total area under coverage is divided into a number of sub areas. Each sub area has its dedicated ceiling mounted transreceiver. Individual VCSEL in the array can be addressed separately and used for transmition of data. In the mobile node a large area photodiode will be used to collect the data from ceiling transmitter. The mobile node will also be mounted with a laser transmitter the output of which will be directed towards the corresponding ceiling receiver for the sub area.

The beam profile of VCSELs being Gaussian in form, to cover a sub area of radius r, the power for each individual laser P is already known by

$$P = ( I_r . \prod . W^2 ) / ( 2 . \exp( -2.r^2 / W^2 ))$$

where $I_r$ is the minimum power density required at the periphery of the sub area. This is determined by the sensitivity of the receiver and the area of the photo detector. The maximum sub area radius for a given lunch power and receiver sensitivity is given by[5]

$$r = \sqrt{ ( P / I_r . \prod .e)}$$

## 3. TRACK AND COMMUNICATE

To achieve mobility in a multiuser environment tracking of the position of a mobile node becomes important in the system. This architecture proposes the use of altrasonic receivers for calculating the position coordinates. An altrasonic transmitter is mounted on the mobile node P and then measure the transmition times of altrasonic energy burst to altrasonic receivers located at points p1, p2 and p3. Speed of altrasonic signal being known, distances from the mobile node to p1, p2 and p3 say d1, d2 and d3 respectively can be easily calculated. This will give equations of three circles of radius d1, d2 and d3 respectively and centers of the circles are p1, p2 and p3 respectively. By subtracting equations from one another and equating with 0 gives equations of three radical axes. Solve of these three equations gives coordinate of the point P (the mobile node).

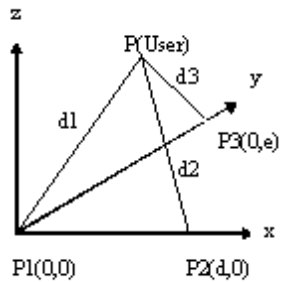

figure 1

The position of the mobile node once being known, the corresponding ceiling transmitter receiver is activated for data communication. To reduce the number of search for mobile nodes a database of the most recent location (sub area/ cell) of each mobile node is maintained. Only when the node is not found in the databases indicated area search operation will be initiated. Some cells/ sub areas in the network is designated as the reporting cells[3]. A mobile node performs a location update only when it enters one of these reporting cells. Whenever a data request arrives the mobile node will be searched in the last reporting cell and the neighboring bounded non- reporting cells. As in the figure whenever a data request for user u arrives, search for user u will be done in the marked positions.

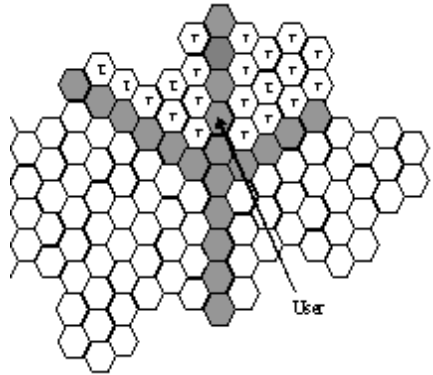

Figure 2. Reporting cells shaded

So in this way tracking becomes much simpler. The ceiling mounted trans-receivers are connected with a backend computer system to allow multiuser operation. Data from a transmitting mobile unit will be directed to the corresponding VCSEL and then through the back end controlling system can be directed to the receiving mobile unit which may be at any positions/ room in the office environment by establishing a link between the corresponding ceiling unit and receiver mobile unit based on the same tracking approach.

**4. SECURED COMMUNICATION**

As more and more data are transmitted over the air the issue of security of the communication becomes an important aspect. The data in the indoor wireless network systems is often of sensitive nature. Any one in the vicinity of the target mobile station can receive the optical signal and interpret the data if no security measure is taken. To ensure that no eaves dropper can breach, the communication between two parties need to be ciphered. A simple authenticated key agreement algorithm based on 'DIFFIE-HELLMAN' technique[1] can be employed. It is assumed that both the users (Alice and Bob) have already agreed on a secret password P before the protocol begins and uses same public values of g and n as the original 'DIFFIE-HELLMAN'. Now Alice and Bob each computes two integers M and ($M^{-1}$) mod (n -1) from the password P. M could be computed in such a way from P so that it gives a unique value and relatively prime with (n-1). Now Alice chooses a random large integer a and sends Bob

$$K1 = g^{(a.M)} \bmod n$$

Bob chooses a random large number b and sends Alice

$$K2 = g^{(b.M)} \bmod n$$

Now Alice computes

$$R = K2^{M^{-1}} \bmod n$$
$$Key1 = R^a \bmod n$$

Bob computes

$$X = K1^{M^{-1}} \bmod n$$
$$Key2 = X^b \bmod n$$

The authenticity of the above scheme can be proved using Fermat's Little Theorem.

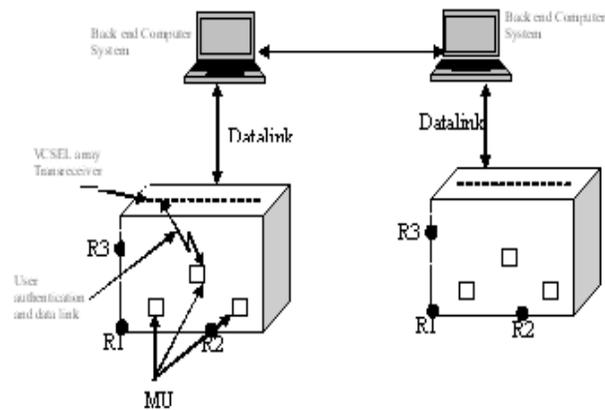

figure 3: Diagrammatic representation of the total system

## 5. CONCLUSION
A novel approach based on intelligent location tracking and management has been suggested. Secured communication is ensured using authenticated key agreement protocol. The proposed architecture is simple and easy to implement.